# A Quantum Description of Radiation Damping and the Free Induction Signal in Magnetic Resonance[†]

James Tropp

General Electric Healthcare Technologies

47697 Westinghouse Drive

Fremont CA, 94539

james.tropp@med.ge.com

## Abstract

We apply the methods of cavity quantum electrodynamics (CQED), to obtain a microscopic and fully quantum-mechanical picture of radiation damping in magnetic resonance, and the nascent formation of the free induction signal. Numerical solution of the Tavis-Cummings model --i.e. multiple spins 1/2 coupled to a lossless single-mode cavity --shows in fine detail the transfer of Zeeman energy, via spin coherence, to excite the cavity -- here represented by a quantized LC resonator. The case of a single spin is also solved analytically. Although the motion of the Bloch vector is non-classical, we nonetheless show that the quantum mechanical Rabi nutation frequency (as enhanced by cavity coupling and stimulated emission) gives realistic estimates of macroscopic signal strength and the radiation damping constant in NMR. We also show how to introduce dissipation: cavity losses by means of a master equation, and relaxation by the phenomenological method of Bloch. The failure to obtain the full Bloch equations (unless semi-classical conditions are imposed on the cavity) is discussed in light of similar issues arising in CQED (and in earlier work in magnetic resonance as well), as are certain problems relative to quantization of the electromagnetic near-field.

## Introduction

Despite longstanding connections between quantum optics and nuclear magnetic resonance [1], NMR theoreticians –excepting those working in force microscopy [2-4] -- have paid scant attention to the Jaynes-Cummings (J-C) model for a two level atom (or a spin ½) coupled to a quantized cavity [5]; much less to that model's extension, by Tavis and Cummings (T-C), to accommodate multiple atoms or spins [6]. Given the centrality of NMR as a tool in modern chemistry, physics, biomedicine, and clinical medicine, it is to be wondered that a fully quantum mechanical theory has not been given, using these comparatively simple tools from cavity quantum electrodynamics (CQED) [7]. This may be due in part to the fact that CQED does not directly yield the Bloch equations, without the assumption of semi-classical conditions, such as a cavity Glauber state [8], or the replacement of quantum mechanical operators by their expectation values. However, earlier approaches [9, 10] to a quantum theory of NMR have not avoided equivalent assumptions; and CQED, without losing rigor, sidesteps other difficulties-- particularly those arising in the application to NMR of near-field quantum electrodynamics, such as quantization of the longitudinal field.

Here we apply the Tavis-Cummings model to NMR. Using the quantized LC circuit [11] as a stand-in for the cavity, we model the time course of Rabi oscillations, for several spins – coherently excited or simply inverted-- which drive an NMR probe initially in its ground state. This allows us to demonstrate -- in a quantum calculation comprising both spins and cavity—not only radiation damping [12-14], but also the nascent formation of the free induction signal. For simplicity we begin with a lossless cavity, and sketch later the inclusion of dissipative processes: cavity losses and spin relaxation.

The theory will be developed along lines which reflect typical experimental practice in CQED, which proceeds in two stages [15] -- first for preparation and then for subsequent evolution and observation. For preparation, the spins are excited by a semi-classical field, which nutates but does not entangle with them; the prepared spins then undergo fully quantum-mechanical evolution, in a low-temperature cavity prepared in a well-defined quantum state of low occupation number, leading to an outcome with spins and cavity entangled. We will assume a preparatory density matrix with spins and cavity all in their ground states, and then apply, *deus ex machina*, a rotation to the spins alone, to produce the desired initial state. The cavity therefore will always start the evolution period in its ground Fock state. The outcomes are assessed by calculating the reduced density matrices for spins and cavity.

NMR practitioners will find some of the results intuitive, and others less so. For example, despite exhibiting the transduction of Zeeman energy into cavity energy-- a prototypical form of radiation damping, even in the lossless cavity, (cf the discussion below under Cavity Damping) , -- the signal exhibits quantum collapse and revival, i.e. decay and recovery through interference of isochromats of incommensurable frequencies [16, 17]. (This is a confounding effect, but unavoidable in these calculations.) Also, driving the cavity with inverted spins gives Rabi oscillation purely of the longitudinal magnetization, with no transverse component developing, i.e. no detectable NMR signal. This reflects the fact that neither the T-C nor J-C Hamiltonian induces a pure rotation of the spins, which, in a loss-free system, would conserve the length of the Bloch vector.

This non-rotational behavior of the quantum Bloch vector is central to the present work, but is not surprising, being pre-figured in the work of J-C [5], and also in the theory of micromasers [18-20]. The classical Bloch equations are not obeyed; nor (equivalently) is the familiar pendulum model, [21-24] which posits that the Bloch angle tracks the displacement angle of a classical pendulum. This relates to another counter-intuitive result, namely that the longitudinal and transverse magnetizations evolve at different frequencies, offset by a factor of two, (with the transverse the slower.) In fact, what is conventionally called the 'Rabi frequency' is that for nutation of the longitudinal moment; and we will adhere to this usage.

As is well known, the optical Bloch equations, which *do* generate rotation of the spins (more properly, of the atomic dipole moment), may be derived from the J-C or T-C Hamiltonian, on the assumption of a cavity excited by a classical oscillatory field, or alternatively, a cavity Glauber (i.e. coherent) state [8, 25-26]. But, while driving the cavity, initially quiescent, with a small number of spin coherences does produce coherences of the field, a fully formed Glauber state is not quick to appear; nor have we seen it in any of our calculations.

We will begin by presenting some elementary theory, and then move on to examples of the microscopic evolution of magnetizations and cavity fields, following suitable excitations. We next introduce dissipative effects: cavity losses and spin relaxation. Then, to trace a path, from a microscopic theory of NMR transduction, to a realistic estimate of the macroscopic signal strength, we calculate the cavity enhancement of Rabi oscillation, using the coupled oscillator model of J-C [5]. This allows us to connect the Rabi nutation frequency with the usual (classical) radiation damping constant, and moves us towards a quantum description of the experimentally observed NMR signal, which has to date remained a work in progress [9-10, 27-31].

**Theory**

The usual presentation of quantum electrodynamics [32] is largely concerned with the radiation field, at the expense of the near-field, for example, that in the vicinity of a nano-scale dipole [33]. The near field ~~need~~ does not radiate; also, it comprises longitudinal as well as transverse components; and its quantization is correspondingly complicated [34-35]. While earlier workers in NMR have sought to address the quantum near-field directly [9], we have chosen here to neglect it, and to focus instead on the cavity operators, which describe the excitation of a resonant LC circuit, or equivalently, a single mode of a tuned cavity. This places our approach squarely in the mainstream of CQED, while allowing a direct focus upon the most consequential process in NMR reception: the transfer of photons from spins to cavity. Due to cavity enhancement of emission -- the cornerstone of CQED [36-37]-- the rate of this process far exceeds that of spontaneous emission by spins into free space.

To write the Hamiltonian we follow the basic procedure of J-C [5], beginning with the details of the NMR antenna, considered as a quantum oscillator, with inductance and capacitance replacing mass and spring constant. We modify the field operators defined by Louisell [11] (with canonical variables electric charge and magnetic flux) by multiplying with $\pm i$ (5), to write the flux (i.e. canonical momentum) in terms of the inductance and Larmor frequency as $\varphi = \sqrt{\hbar\omega_0 L/2}[\hat{a}+\hat{a}^\dagger]$. Postulating (for definiteness) that the inductor comprises a singly wound Helmholtz pair, we approximate the laboratory frame radiofrequency field $B_1$ as $\varphi/2a$, where $a$ is the window aperture in meters-squared. This leads directly to the Rabi fundamental frequency, i.e. $\Omega_0 = \gamma\sqrt{\hbar\omega_0 L/2}/2a$ where $\gamma$ is the gyromagnetic ratio, and the factor of ½, required for the correct nutation rate, is implicit in the partition of field operators according to the rotating wave approximation.

Then for the coupling of several spins to a single cavity mode we write the Tavis-Cummings (i.e. extended Jaynes-Cummings) Hamiltonian in the interaction picture:

$$\mathfrak{H} = -\hbar\frac{\Omega_0}{2}\sum_{j=1}\{\hat{a}^\dagger \hat{I}_+^{(j)} + \hat{a}\hat{I}_-^{(j)}\} \quad [1]$$

where the Rabi fundamental $\Omega_0$ is effectively the coupling constant between spins and cavity, and the frequency for the transition connecting the Fock states $|n\rangle$ and $|n+1\rangle$ is $\Omega_n = \sqrt{(n+1)}\Omega_0$; (for large $n$ we will approximate $\sqrt{n+1}$ with $\sqrt{n}$.) The sum is over spins; the operator pairings are for a spin ½ with positive gyromagnetic ratio. The Zeeman interaction is implied by the presence of the Larmor frequency (above), but a direct treatment of the static polarizing field is unnecessary.

For the case of a single spin (the J-C model), several informative results are obtainable by elementary analytic calculation. Starting from the preparatory density matrix $\rho^{(prep)} = |0\alpha\rangle\langle 0\alpha|$, and applying (as described earlier) a spin rotation of $\pi/2$, (about the *y* axis of a rotating reference frame) we arrive the density matrix describing our initial conditions:

$$\rho^{(init)} = 1/2 \begin{bmatrix} 1 & -1 & 0 \\ -1 & 1 & 0 \\ 0 & 0 & 0 \end{bmatrix}, \quad [2]$$

where the product basis elements $|0\alpha\rangle, |0\beta\rangle, |1\alpha\rangle$ give the occupation numbers and spin projections. Then with the abbreviations $c = \cos \tfrac{1}{2}\Omega_0 t$ and $s = \sin \tfrac{1}{2}\Omega_0 t$, the density matrix, at time *t*, evolves, to:

$$\rho(t) = 1/2 \begin{bmatrix} 1 & -c & is \\ -c & c^2 & -isc \\ -is & isc & s^2 \end{bmatrix}. \quad [3]$$

The reduced [38] spin and photon density matrices are:

$$\rho^{(spin)}(t) = 1/2 \begin{bmatrix} 1+s^2 & -c \\ -c & c^2 \end{bmatrix}. \quad [4]$$

and

$$\rho^{(cavity)}(t) = 1/2 \begin{bmatrix} 1+c^2 & is \\ -is & s^2 \end{bmatrix}. \quad [5]$$

From these expressions it is easily seen that the longitudinal magnetization oscillates at the Rabi fundamental frequency, and the transverse at one half that value; also the cavity one-quantum coherence and transverse magnetization are in time quadrature in the laboratory frame (the two are related like the real and imaginary components of the complex NMR signal.) The case of an initial nutation of $\pi$ is easily worked out, and shows a perfect absence of transverse magnetization. These effects will be borne out in the numerical examples presented below.

## The Case of Two Spins

As above, the basis set comprises simple product kets, e.g. $|n\alpha\beta\rangle$, giving the Fock state indices (0, 1, and 2) and the spin projections. The eigenvalues, in units of $\Omega_0$, are $\pm\sqrt{3/2}$, $\pm 1$, $\pm\sqrt{1/2}$, and 0 (multiplicity 6). The twelve members of the basis set include seven elements, which would together comprise a pair invariant subspaces of constant excitation 1 and 2 [39]; but a redundant basis is easily adaptable to the case of more spins and is therefore preferred.

Excited states of the spins are generated, as noted, by applying spin rotations ($\pi/2$ or $\pi$) to the preparatory density matrix $\rho^{(prep)} = |0\alpha\alpha\rangle\langle 0\alpha\alpha|$; also as noted, the rotation is performed by a classical field; the subsequent time evolution of the spins, and their prompt entanglement with the cavity, is then determined by solution of the Liouville equation, based upon the Hamiltonian of Eq [1]. The form of $\rho^{(prep)}$ is appropriate for small numbers of spins, and also ensures a trace of unity. Since the cavity is presumed lossless, and the NMR linewidth perfectly homogeneous, excitation along the rotating axes *x* or *y* ensures that all off-diagonal elements of the reduced spin density matrix, $\rho^{(spin)}$, are either pure real or pure imaginary.

Figure 1A shows the non-classical time evolution of the longitudinal and transverse magnetizations (for two periods of the Rabi fundamental), starting from an initial condition with both spins tipped by $\pi/2$ and the cavity is in its ground state. The subsequent time course is calculated by evolving the propagator in the eigenbasis, following numerical diagonalization of the Hamiltonian [40]. The magnetizations are plotted at baseband, i.e. with the harmonic time dependence at the Larmor (carrier) frequency having been removed by demodulation. All possible spin coherences are excited, of orders zero, one, and two, although only the latter two are visualized; the dotted trace shows the total number of excitations, constant and equal to 1.

The longitudinal magnetization (blue trace) oscillates at the Rabi frequency. The transverse magnetization (a single-quantum coherence) oscillates at half the Rabi frequency, $\sqrt{1/2}\Omega_0$, between the negative and positive *x* axes of the rotating coordinate frame [41]. Since the two magnetizations evolve at different frequencies, they cannot be said to be in time quadrature; nonetheless, the extrema of one occur at or near the zeros of the other, as expected for the time evolution of the corresponding classical signals. The incipient damping, of both magnetizations, signals the onset of quantum collapse [17]; revival is easily demonstrated in a time course of longer duration.

The red trace (shown at doubled amplitude for better visualization) is the two-quantum coherence (a Schrödinger's cat state), It oscillates (roughly sinusoidally) at the highest eigenfrequency, $\sqrt{3/2}\Omega_0$.

Figure 1B shows the time evolution of the cavity coherences (also at baseband.) There are two single quantum coherences (blue and green traces) with complicated time dependences, representing the summed cavity reduced density matrix elements $\rho_{12}^{(cavity)} \pm \rho_{21}^{(cavity)}$ and $\rho_{23}^{(cavity)} \pm \rho_{32}^{(cavity)}$, where the index 1 denotes the ground Fock state. The red trace undergoes approximately sinusoidal oscillation at $\sqrt{3/2}\Omega_0$, and represents the two-quantum coherence $\rho_{13}^{(cavity)} \pm \rho_{31}^{(cavity)}$; its time evolution tracks that of the corresponding spin coherence. The phase of the one quantum coherences depends upon the choice of axis for the initial excitation of the spins; for initial rotation about $y$ (leading to purely real *spin* coherences, i.e. entirely expressed in terms of operator products, $I_x^{(i)}$ and $I_z^{(j)}$), the single-quantum *cavity* coherences are pure imaginary, of which more below.

Figure 2A illustrates our central result: that is, the transfer of Zeeman energy (via spin coherence) from the precessing spins to the tuned cavity, with the concomitant appearance of an induced field. The transverse magnetization (cf Fig. 1A) is shown in green for reference.

Then summation of the pair of one-quantum cavity coherences (cf Fig. 1B) yields the quasi-sinusoidal waveform shown in dashed blue. Fourier analyses (not shown) demonstrate minor differences in frequency content between the transverse magnetization and the summed cavity coherences, despite which the two are approximately in time quadrature at baseband, as is expected [13] in a conventional pulsed NMR experiment.

The dashed black trace gives the expectation value at baseband of magnetic flux (normalized to the square root of the occupation number). Since the off-diagonal terms of the cavity reduced matrix are here purely imaginary, its trace with the summed field operators vanishes when the operator time dependence (at the Larmor frequency) is omitted. The summed operators may nonetheless be evaluated at Larmor, yet plotted at baseband, i.e. the phasor amplitude plotted, as we have done. Voltage and current are in phase for a perfectly tuned oscillator excited at resonance, and the net amplitude of flux – or field-- scales directly with current.

The time course of flux directly tracks that of the induced magnetic field. Since the transverse moment oscillates at about one half the Rabi frequency, and since our cavity is assumed lossless, the zero of transverse moment coincides with the maximum field, in contrast to a classical model with losses (which is discussed later.) Overall, the figure illustrates the dynamics of NMR transduction, which may be described qualitatively as the conversion of spin coherence to cavity coherence. It may also be viewed as the primordial form of radiation damping -- even in the absence of cavity losses-- since Zeeman energy is transduced to create an oscillatory field. (We return to this point below, in the section on cavity enhancement.)

For reference, Figure 2B shows the phenomena of collapse and revival –here of longitudinal magnetization -- observed over several periods of the Rabi fundamental, following preparation with a nutation pulse of $\pi/2$ ( above). After a $\pi$ pulse (below), pure Rabi oscillation is observed without quantum collapse.

## The Case of Multiple Spins ($N > 2$)

We show in this section some details of dynamic behavior for larger numbers of spins; we restrict consideration to an initial density matrix $\rho^{(prep)} = |0\alpha...\alpha\rangle\langle 0\alpha...\alpha|$. Figure 3A shows the time evolution of transverse magnetization, over a single period of the Rabi fundamental, for spin clusters with $N$ ranging from two to seven, and the vertical scale normalized correspondingly. The curves show the expected behavior, inasmuch as the decay rate of transverse magnetization (which essentially equals the emission rate in the semi-classical regime) grows with the number of emitted photons, and therefore demonstrates stimulated emission. This is clearly seen in the figure by tracing the position of the first zero crossing, for progressively larger $N$. The longitudinal magnetization (not shown) displays similar behavior, although not as pronounced.

Typical cavity dynamics are complex, and are shown in Figure 3B, which gives, over the same time interval, (in solid lines) the time courses of the individual one-quantum cavity coherences, for five spins, – each calculated as above by summing conjugate elements), plus (in dashed blue) their net resultant, which exhibits the expected quasi-sinusoidal shape (cf Figure 2A), also with evidence of quantum collapse. The net magnetic flux (dashed black, calculated as describe earlier) follows closely the total summed coherences. The rapid initial growth of cavity one-quantum coherence is explained by the early dominance of the 1,2 element (plus conjugate) in the cavity density matrix, which involves the cavity ground state – fully populated at the outset. The other tributaries (which we write in abbreviated form) -- e.g. $\rho_{23}^{(cavity)}$, $\rho_{34}^{(cavity)}$, $\rho_{45}^{(cavity)}$ – show the expected delayed growth characteristic of stimulated emission, as higher cavity levels are progressively populated.

Fig. 4A shows, for five spins in a single period of the Rabi fundamental, the interchange of the net photon population (dotted blue trace) with transverse (green) and longitudinal (blue) magnetizations. For comparison, we show in Figure 4B the analytic results (starting from a negative transverse moment) for a single spin (above) for two periods of the Rabi fundamental. Note the absence of quantum collapse and resulting high symmetry. Otherwise these are similar to the results of Figure 4A, and show usefulness as a qualitative guide.

**Introduction of Cavity Losses and Relaxation**

Since a complete theory of radiation damping will include dissipation, we sketch the introduction of cavity losses and spin relaxation. The examples are illustrative, not definitive. Earlier attempts at a quantum theory of the NMR signal treated cavity damping artificially [9-10, 42], e.g. by simple escape of photons from an active volume, or by a transmission line dashpot , (i.e. a non reflective termination.) Here we employ the master equation from the theory of micromasers [18-20], which derives from coupling to a thermal bath, and gives the damping contribution for $\rho$:

$$\dot{\rho} = (\gamma/2)\{(n+1)(2\hat{a}\rho\hat{a}^{\dagger} - \hat{a}^{\dagger}\hat{a}\rho - \rho\hat{a}^{\dagger}\hat{a}) + n(2\hat{a}^{\dagger}\rho\hat{a} - \hat{a}\hat{a}^{\dagger}\rho - \rho\hat{a}\hat{a}^{\dagger})\},$$

[6]

where $n$ is the mean photon occupation number, and $\gamma$ is now the photon damping rate. For micromasers, $\rho$ is usually the reduced photon density matrix [18-19], but here it is the combined spin-photon matrix. The dissipative term in Eq. 4 is added to the coherent Liouville equation, to give the master equation with damping. To this we then superadd the effects of relaxation, following the phenomenological Bloch equations, with separate decay constants for the longitudinal and transverse moments; for simplicity both are assumed to damp to zero. We consider a single spin, since multispin longitudinal order associated with high polarization [43] requires additional magnetization modes and decay constants [44]. The full master equation is then solved numerically by iteration, using the Euler method (45).

Figure 5 shows that the effects of cavity losses and spin relaxation can be adjusted empirically to mimic the results of classical radiation damping in NMR, that is, to produce substantial damping of Rabi oscillation in half a cycle. Figure 5A gives the damped signal from a single spin; figure 5B shows a realistic calculation based upon the classical Bloch-Kirchhoff [12, 46] equations, for a sample of water protons at 14 tesla, using coil and sample parameters derived from an experimental study [14]. Of particular interest is the peak damping current, which reaches a value of 7 milliamps. The calculated damping linewidth here -- based upon the measured coil efficiency, and excluding the (unmeasureable) filling factor-- is 55 Hz, compared with a measured value of 65 Hz. Also, compare particularly Fig. 5A with Figure 4B (one spin, no dissipation, no quantum collapse), to judge the rapidity of signal decay with losses. The magnetizations have similar trajectories in both figures 5A and 5B; also the oscillator current in 5B resembles the photon population in 5A, when account is taken of the early time behavior (inset) .

**Cavity Enhancement and Stimulated Emission in NMR: the Rabi Frequency and the Radiation Damping Constant**

We first numerically estimate the cavity-enhanced Rabi fundamental; we then establish a relationship between the Rabi frequency and the radiation damping constant; we also discuss the importance of stimulated emission in setting the signal power in NMR (i.e. the rate of energy transfer from spins to cavity). Throughout, we follow the coupled oscillator model of J-C [5], i.e. for a radiating atom coupled to a single cavity mode, or in our case, a quantum oscillator. For a reasonable Helmholtz coil, (round windows of inside diameter 0.75 cm, separated by the diameter, inductance of 58 nH) the Rabi fundamental (cf Theory) takes the value $\Omega_0 = 8.13 \times 10^{-5} \sec^{-1}$ for a proton in a polarizing field of 14.1 tesla, i.e. with Larmor frequency $\omega_0/2\pi$ of 600 MHz. This corresponds to an emission time (1/2 of the Rabi period) of $6.47 \times 10^{-6}$ sec. By contrast, the inverse lifetime for spontaneous emission at 600 MHz, by a single excited proton in free space, is: $\mu_0 \hbar \gamma^2 \omega^3 / \pi c^3 = 6 \times 10^{-21} \sec^{-1}$. The staggering difference (~15 orders of magnitude) is due to the high concentration of magnetic flux created per photon, inside the coil, relative to that in free space. J-C estimate a cavity enhancement of $\sim 10^7$ for spontaneous emission in the ammonia beam maser [5].

Then the *rotating* frame $B_1$ field for our model NMR coil carrying unit current is just $L/4a$. Writing the current in terms of the oscillator mean occupation number $n$ as $\sqrt{2n\hbar\omega_0/L}$, (according to $\frac{1}{2}LI^2 = n\hbar\omega_0$), and multiplying by field per current and the gyromagnetic ratio, yields $\gamma\sqrt{n\hbar\omega_0 L/2}/2a$, which, for large $n$, is just the Rabi nutation frequency connecting the $n$tt and $n+1^{st}$ Fock states. This is therefore the product of a nutation rate per unit current, and a current. Incidentally, this can also be written in terms of the occupation number $n$ as $\Omega_n = \Omega_0 \sqrt{n}$, which shows the importance of stimulated emission in setting the overall emission rate in NMR; we shall return to this point below.

The classical radiation damping constant [12] also factors as the product of a nutation rate per current, and a current. We start by writing it in terms of the fill factor $\eta$, as $k = \mu_0 \gamma \eta M_0 Q / 2$ (in SI units), where $Q$ is the source-loaded tuned-circuit quality factor, $M_0$ is the equilibrium magnetization, $\gamma$ is the gyromagnetic ratio and other symbols have their usual meanings. Using reciprocity theory [14, 47-48], this may be rewritten as $k = \gamma \omega_0 V M_0 \zeta^2 / 4$ with the transceiver efficiency defined as $\zeta = B_1(I) / (I \sqrt{R})$, where $B_1(I)$ is the laboratory frame radiofrequency field at coil current $I$, and $R$ is the coil resistance, (with*out* source loading); $V$ the sample volume. Thus the damping constant is a product of two factors: $\gamma B_1(I)/2I$ and $\omega_0 V M_0 B_1(I)/2IR$: the first a nutation rate per unit current, and the second (per reciprocity theory) a current, i.e. voltage over resistance. This establishes (if it were doubted) the close relationship of Rabi nutation to radiation damping. Even for a lossless cavity, we have used the term 'radiation damping' to describe the diminution of the transverse moment, and concomitant growth of the longitudinal, (as the spins emit), by analogy with the Wigner-Weisskopf theory of spectroscopic linewidth [11], in which the fundamental event is the emission of a photon, whose subsequent fate (e.g absorption by a black body) is of small interest.

Finally we return to the question of stimulated emission, specifically for the water protons in a realistic NMR probe as given elsewhere, [14] and described in the legend of Fig. 5B. From the familiar equation for Zeeman energy balance in terms of nutation angle ($\dot{E} = M_0 V B_0 \dot{\vartheta} \sin \vartheta$) we calculate, for $\dot{\vartheta} = \Omega_0$, a net power due to coherent spontaneous emission of 11 $pW$, that is, for nutation of the net moment at the Rabi fundamental, without stimulated emission. This is far below the rms power of 25 $\mu W$, calculated classically from Bloch-Kirchhoff with a peak oscillator current of 7 mA (cf Figure 5B above), and the given resistance (with source loading) of 1.0 ohm [14]. However, the oscillator occupation number for 7 mA is $3.57 \times 10^{12}$, yielding an increase in the Rabi frequency of a factor $\sqrt{n} = 1.9 \times 10^6$. The resultant emitted power is now increased from 11 $pW$ to 21 $\mu W$, which lies within 20% of the 25 $\mu W$ calculated classically. (In reckoning the classical power, the amplitude of the current must be treated as AC, and its rms value used. This is verified by a direct calculation in terms of the Zeeman energy, as gotten from the longitudinal moment.)

The numerical agreement between the two values of power (21 $\mu W$ and 25 $\mu W$) is perhaps fortuitous, given the approximate nature of the calculations-- but it is nonetheless consistent with the view that the power in radiation damping has a substantial contribution, amounting to several orders of magnitude, from stimulated emission. This result is also is also consistent with the increase of the Rabi frequency measured in experiments on large populations of excited Rydberg atoms in a tuned cavity [49], as well for single atoms in the presence of larger injected fields [50].

## Discussion

Cavity quantum electrodynamics originated with the observation by Purcell, that spontaneous emission inside a tuned cavity is enhanced by the increased the density of states [36-37]. Bloembergen and Pound [12] then argued that the observed signal power in NMR results from the twinned factors of cavity enhancement and coherent spontaneous emission; and this viewpoint has been accepted for decades [30].

Our own calculations suggest that these two factors do not suffice to explain the strength of the FID signal, but that stimulated emission makes an essential contribution, amounting to several orders of magnitude. The treatment of the enhanced radiation density inside the cavity then becomes important, as we follow here not Bloembergen-Pound, but Jaynes-Cummings [5].

That is, B-P, -- and others [51-52] -- typically write the enhancement factor (i.e. the density of states) in terms of the cavity $Q$, as a measure of the sharpness of the cavity resonance. This approach blurs the distinction between the atom-field coupling constant and the cavity dissipation rate, as set out in the bad cavity limit [53], and also as manifest in the master equation, which clearly separates the coherent atom-field interaction from the incoherent cavity damping.

J-C, in contrast to B-P, write the Hamiltonian directly in terms of the atom-field coupling constant, without reference to cavity dissipation. This also comports with the theory of reciprocity [47-48], inasmuch as the emf in nuclear magnetism depends only upon the radiofrequency B field per unit oscillator current, which determines the nutation rate (i.e. Rabi frequency) given the actual current, and serves in the classical theory as an analog to the coupling constant between spins and cavity. The J-C treatment is also intrinsically a theory of large nutations, in contrast to that of B-P, which (despite solving the *classical* damping equation for large excursions of the Bloch vector) relies nonetheless on small perturbations for the *quantum* treatment of cavity-enhanced emission, with the density of states entering through the Fermi Golden Rule.

On a different tack, we have noted earlier that CQED does not lead directly to the Bloch equations, without strong assumptions amounting to the imposition of semi-classical behavior of the cavity; this path has been chosen by other workers in NMR [9-10] including an instance in which the validity of Bloch equations is taken to define the zero order condition in a perturbation scheme [9]. In a larger sense, the question of the transition to semi-classical behavior remains open in CQED [21, 54-56]. The disappearance of quantum collapse and revival has been proposed as a marker for the arrival of classical or semi-classical behavior [54] ; but in practical NMR, where low resonator quality factors (~50 to 500) enforce the bad cavity limit [53], collapse and revival have never been (and likely never will be) observed. We therefore propose that the transition to classical behavior in NMR comes with the onset of rotation of the Bloch vector, which we take to coincide with the appearance of a cavity Glauber state. That is, if the Liouville equation for a spin ½ coupled to a cavity be written for the initial cavity state a Glauber state, then the Liouville matrix elements are diagonal in photon variables, and pure rotations of the Bloch vector can occur. On the other hand, if the cavity starts in a Fock state, the Hamiltonian disallows pure rotations of the spins, and non-classical effects ensue-- an issue which has not been addressed by prior workers in magnetic resonance [9-10, 27-31]. The theory given here, although incomplete, demonstrates the gap between the pure quantum and semi-classical regimes, with the latter corresponding to the customary world of NMR observations.

Bridging the gap between the two will probably require calculations including both dissipation and many more spins.

## Acknowledgement

This work was supported by General Electric Healthcare Technologies.

**Figure Legends**

Figure 1. Time evolution of magnetizations and cavity coherences. For two periods of the Rabi fundamental, following a $\pi/2$ pulse of two perfectly polarized spins. Spin and cavity coherences are offset in time by one quarter Larmor period. A) Longitudinal magnetization (blue), transverse magnetization (green), two quantum coherence (red.). Vertical axis normalized to number of spins. The dotted black line is the number of excitations, constant at 1.0. B) Individual one-quantum cavity coherences: $\rho_{12}^{(cavity)} \pm \rho_{21}^{(cavity)}$ (blue) and $\rho_{23}^{(cavity)} \pm \rho_{32}^{(cavity)}$ (green); and also two quantum cavity coherence $\rho_{13}^{(cavity)} \pm \rho_{31}^{(cavity)}$ (red).

Figure 2. Energy transfer from spins to cavity, and Rabi oscillation of longitudinal moment. A), Time course of transverse magnetization (green), and summed one quantum cavity coherences (dotted blue.) Also, (dotted black) the normalized magnetic flux (per square root photon). B) Time

course of longitudinal magnetization following preparatory nutation of $\pi/2$ (above, note quantum collapse and revival) and $\pi$ (below), for 10 periods of the Rabi fundamental. *Note the division of the vertical scale for the upper and lower traces*.

Figure 3. A) Illustration of stimulated emission over one period of the Rabi fundamental: time courses of transverse magnetization for increasing numbers of spins, from two – (solid blue trace) through seven (dotted red trace.) (Number of spins increases in color sequence blue, green, red, and from solid to dotted trace) The damping (emission) rate is indicated by the first zero crossing, which arrives progressively sooner with each additional spin, as expected from stimulated emission. Vertical axis normalized to number of spins. B) One quantum cavity coherences (solid traces) for five spins, and their summations, to give approximately sinusoidal resultant (dashed navy trace); also, weighted sum (dashed black trace) gives net flux. Color sequence of one quantum coherences (solid lines) starting from $\rho_{12}^{(cavity)}$ is navy, red, green, violet, cerulean. Note the acceleration relative to two spins shown in Figure 2. Refer to text for details.

Figure 4. Comparison of numerical and analytic results. A) Signal formation with five spins, following nutation of $\pi/2$, for one period of the Rabi fundamental. Longitudinal moment blue, transverse green, total photons dashed blue, total excitations dashed black. B) Analytical solution of Jaynes-Cummings model, single spin in cavity tuned at Larmor

frequency, following nutation of $\pi/2$, for *two* periods of the Rabi fundamental. Colors as in 3A.

Figure 5. Effects of dissipation: cavity losses and spin relaxation. A) J-C model (one spin) with cavity damping and spin relaxation, to simulate radiation damping in conventional NMR. Strong cavity damping ($\gamma = 5/\tau_0$), short $T_2$, ($\tau_0/4$) long $T_1$ ($\tau_0/0.1$), with $\tau_0$ the Rabi fundamental period. Longitudinal and transverse magnetization solid blue and green respectively, total photons dotted blue, total excitations dotted black. B) Classical radiation damping per the Bloch Kirchhoff equations. Magnetizations as in 5A, magnitude current in dotted blue (milliamp scale). Inset: early time current, 1 µs duration, 8 mA excursion starting at zero; compare zero initial photons in 4A. Details of the sample and the measurement of the coil efficiency match *experimental* conditions [14]. Starting magnetic moment: $9.73 \times 10^{-9}$ amp-meter$^2$, corresponds to neat water at T = 298 K in a 5 mm NMR tube, with vertical probe window of 1.6 cm. Measured RF coil efficiency: $2.64 \times 10^{-4}$ tesla/$\sqrt{\text{watt}}$. Measured quality factor of $Q = 220$, plus assumed coil resistance of $R = 1.0$ ohm (both $Q$ and resistance are source *loaded*), gives coil inductance $L = 58$ nH.

Tropp Figure 1

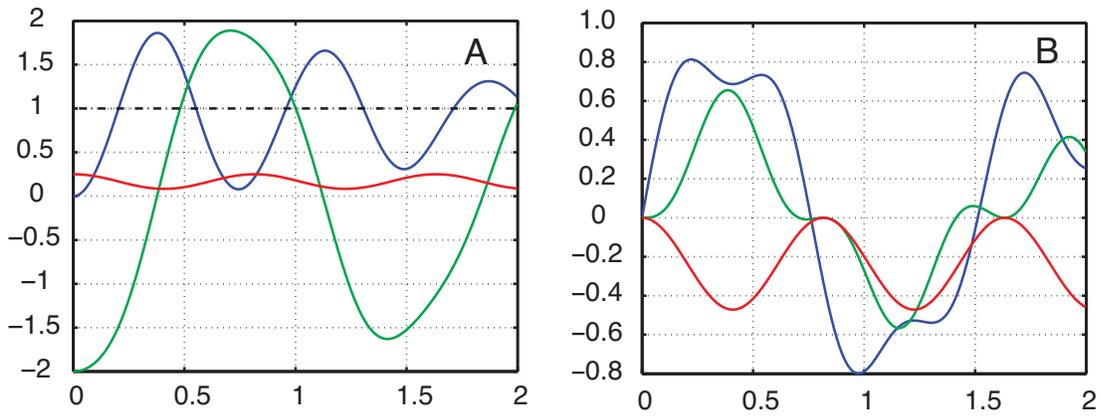

Tropp Figure 2

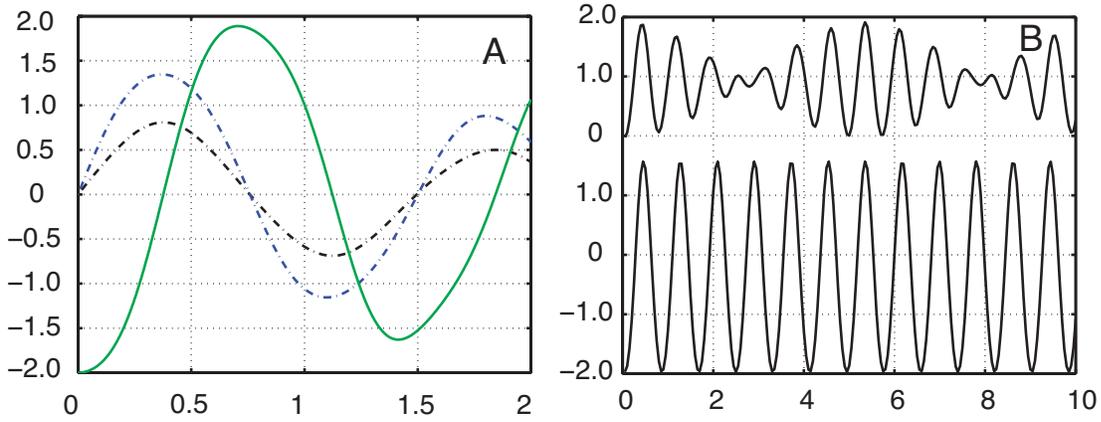

Tropp Figure 3

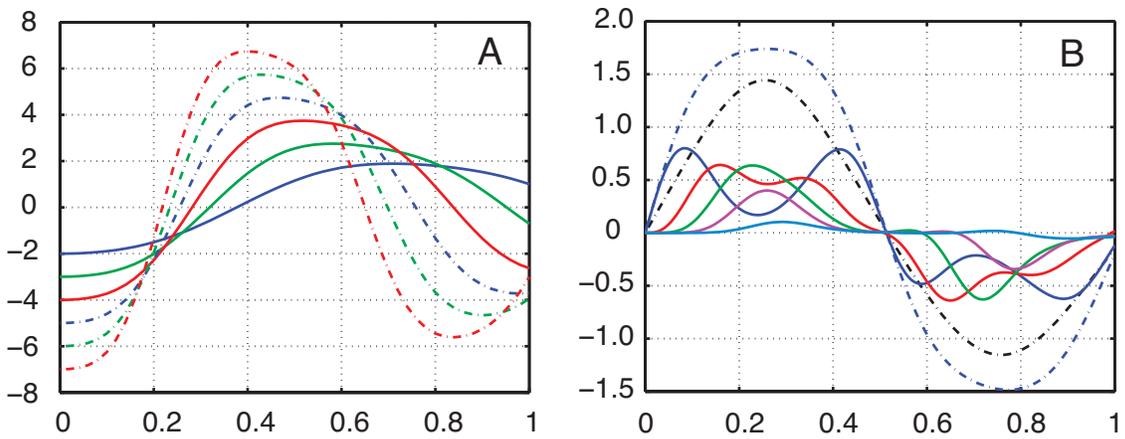

Tropp Figure 4

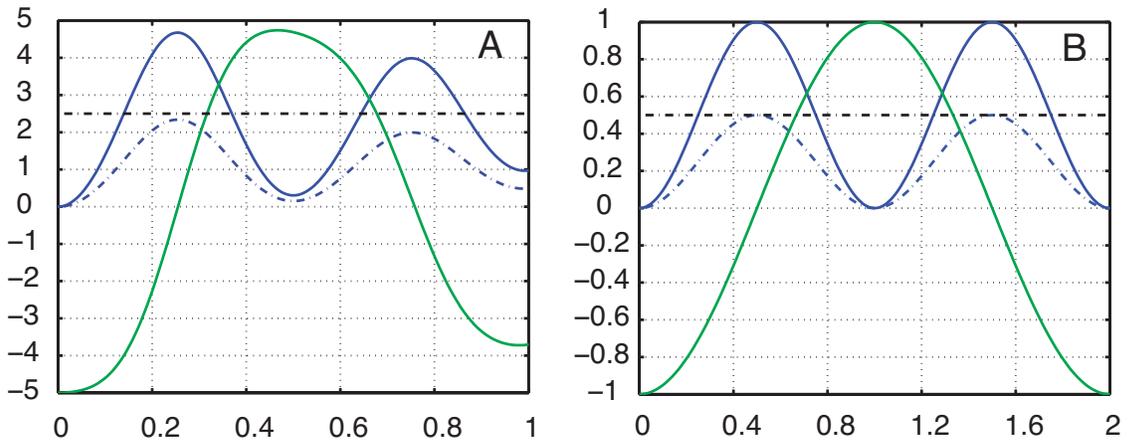

Tropp Figure 5

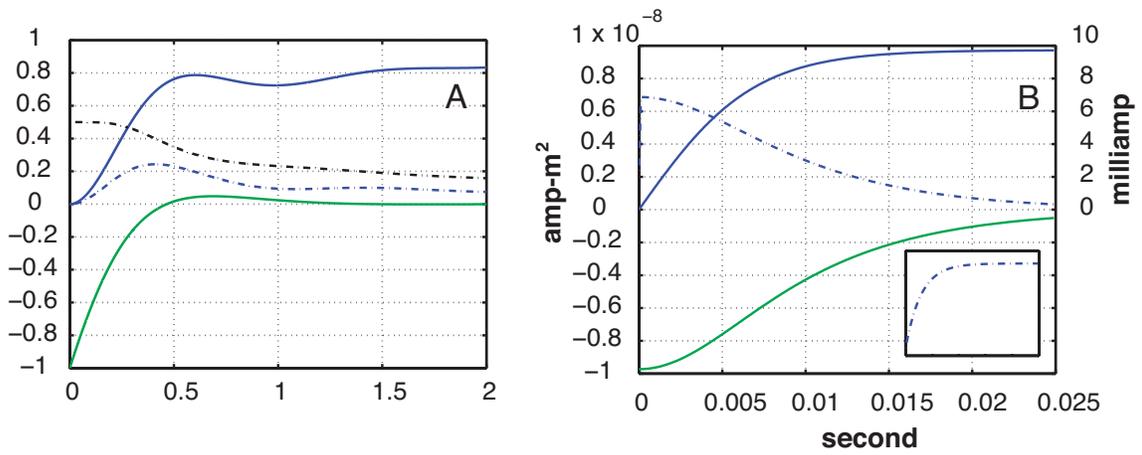

31